\documentclass{tMOP2e}

\usepackage{subfigure}
\usepackage{color}

\theoremstyle{plain}

\theoremstyle{definition}

\theoremstyle{remark}

\begin{document}


\articletype{ }

\title{Nonlinear interferometer for tailoring the frequency spectrum of bright squeezed vacuum}

\author{T.~Sh.~Iskhakov$^{\rm a}$$^{\dag}$\thanks{$^\dag$ Present address: Technical University of Denmark Department of Physics, Fysikvej 309, 2800 Kgs. Lyngby, Denmark\vspace{6pt}}, S.~Lemieux$^{\rm b}$, A.~Perez$^{\rm a,c}$, R.~W.~Boyd$^{\rm b,d,e}$, G.~Leuchs$^{\rm a,b,c}$, and M.~V.~Chekhova$^{\rm a,c}$$^{\ast}$\thanks{$^\ast$Corresponding author. Email: maria.chekhova@mpl.mpg.de
\vspace{6pt}}
\\\vspace{6pt}  $^{a}${\em{Max-Planck Institute for the Science of Light, Guenther-Scharowsky-Str. 1 / Bau 24, Erlangen  D-91058, Germany}};\\
$^{b}${\em{Department of Physics, University of Ottawa, Ottawa, ON, Canada K1N 6N5}};\\
$^{c}${\em{University of Erlangen-N\"urnberg, Staudtstrasse 7/B2, 91058 Erlangen, Germany}};\\
$^{d}${\em{Institute of Optics and Department of Physics and Astronomy, University of Rochester, Rochester, NY 14627 USA}};\\
$^{e}${\em{School of Physics and Astronomy, University of Glasgow, Glasgow G12 8QQ,UK}}\\
\received{v4.1 released September 2014} }

\maketitle

\begin{abstract}
We propose a method for tailoring the frequency spectrum of bright squeezed vacuum by generating it in a nonlinear interferometer, consisting of two down-converting nonlinear crystals separated by a dispersive medium. Due to a faster dispersive spreading of higher-order Schmidt modes, the spectral width of the radiation at the output is reduced as the length of the dispersive medium is increased. Preliminary results show 30\% spectral narrowing.


\end{abstract}

\section{Introduction}

Bright squeezed vacuum (BSV) is a state of light emerging from the output of a high-gain unseeded parametric amplifier (OPA). Due to its nonclassical properties such as photon-number entanglement and quadrature squeezing, this state  is useful for various quantum-information applications, among them quantum metrology~\cite{metrology}, quantum imaging~\cite{q-imaging}, and quantum lithography~\cite{lithography}. Besides containing a high number of photons in each mode, the state is essentially multimode, both in the frequency and in the angle. These features provide its large information capacity, as quantum information can be encoded in the number of photons in different modes. At the same time, the presence of a large number of modes can be a disadvantage in certain experiments, for instance achieving phase super-sensitivity~\cite{super} or (related) gravitational-wave detection~\cite{gravit}. A possible way to reduce the number of modes without losing nonclassical correlations is to use a nonlinear interferometer, in which only part of the spectrum is amplified. This has been already demonstrated for the angular spectrum in Ref.~\cite{separation}. The goal of this work is to show similar behavior in the frequency domain.

The paper is organized as follows. In the next two subsections, we briefly describe the mode structure of BSV (subsection~\ref{S1.1}) and the idea and operation of a nonlinear interferometer (subsection~\ref{S1.2}). Section~\ref{S2} explains the idea of reducing the number of modes in BSV in space/angle and time/frequency and demonstrates the narrowing of the BSV angular spectrum in a nonlinear interferometer with spatially separated crystals. The experiment on the narrowing of the BSV frequency spectrum is described in Section~\ref{S3}. Section~\ref{S4} contains the conclusions.

\subsection{BSV and its eigenmodes}\label{S1.1}

The most convenient way of generating BSV is high-gain parametric down-conversion in a nonlinear crystal, which can be considered as an unseeded traveling-wave OPA. The frequency-angular spectrum and photon-number correlations are well described by the Bloch-Messiah formalism~\cite{Wasilewski,Silberhorn,Bloch}, in which the Hamiltonian is diagonalized by passing to the eigenmodes of the OPA. For instance, in the case of spatially multimode PDC, the Hamiltonian can be written as~\cite{Bloch}
\begin{equation}
H=i\hbar\Gamma\iint d \mathbf{q}_s  d \mathbf{q}_i F(\mathbf{q}_s,\mathbf{q}_i)a^\dagger_{\mathbf{q}_s}a^\dagger_{\mathbf{q}_i}+h.c.,
\label{eq:Ham}
\end{equation}
where $\Gamma$ characterizes the coupling strength, $\mathbf{q}_{s,i}$ are the transverse wavevectors of the signal and idler radiation, and $a^\dagger_{\mathbf{q}_{s,i}}$ are the photon creation operators in the corresponding plane-wave modes. The central part of the Hamiltonian is the two-photon amplitude (TPA), $F(\mathbf{q}_s,\mathbf{q}_i)$, whose meaning is the probability amplitude of a photon pair created with the wavevectors $\mathbf{q}_s,\mathbf{q}_i$. The Hamiltonian (\ref{eq:Ham}) is diagonalized by representing the TPA as a Schmidt decomposition,
\begin{equation}
F(\mathbf{q}_s,\mathbf{q}_i)=\sum_k \sqrt{\lambda_k}u_k(\mathbf{q}_s)v_k(\mathbf{q}_i),
\label{eq:TPA Schmidt}
\end{equation}
where $\lambda_k$ are the Schmidt eigenvalues, $u_k(\mathbf{q}_s),\,v_k(\mathbf{q}_i)$ the Schmidt modes, and $k$ is a two-dimensional index. By definition, the modes are ordered so that $\lambda_{k+1}\le\lambda_k$. The Hamiltonian (\ref{eq:Ham}) can be now written as a sum of two-mode Hamiltonians,
\begin{equation}
H=\sum_k \sqrt{\lambda_k}H_k,\,\,H_k=i\hbar\Gamma A^\dagger_kB^\dagger_k+h.c.,
\label{eq:BM_dec}
\end{equation}
with the photon creation operators $A^\dagger_k,\,B^\dagger_k$ relating to the Schmidt modes. Moreover, if the signal and idler beams are indistinguishable, their Schmidt modes are the same, and
\begin{equation}
H_k=i\hbar\Gamma (A^\dagger_k)^2+h.c.
\label{eq:Ham_deg}
\end{equation}
It is worth mentioning that the Schmidt decomposition can be alternatively performed in the space coordinates, which is equivalent to the wavevector decomposition~(\ref{eq:TPA Schmidt}). A similar decomposition is valid for the frequency/temporal domain. In different works, the Schmidt modes are also called nonmonochromatic modes~\cite{Opatrny}, squeezing (eigen)modes~\cite{Boyd}, broadband modes~\cite{Silberhorn}, or supermodes~\cite{Fabre}.

Clearly, in terms of these new modes, photon-number correlations are only pairwise. The total mean photon number can be represented as a sum of incoherent contributions from all Schmidt modes,
\begin{equation}
\langle N\rangle=\sum_k \langle N_k\rangle,\,\,\langle N_k\rangle=\sinh^2[\sqrt{\lambda_k}G],
\label{eq:ph_N0}
\end{equation}
where $G=\int\Gamma dt$ is the parametric gain. This means that while at low gain ($G<<1$), the Schmidt modes are populated with the weights given by the Schmidt eigenvalues $\lambda_k$, at high gain these weights are changed to become~\cite{Bloch}
\begin{equation}
 \tilde{\lambda}_k=\frac{\sinh^2[\sqrt{\lambda_k}G]}{\sum_k\sinh^2[\sqrt{\lambda_k}G]}.
\label{eq:renorm}
\end{equation}
According to this, at high gain the lower-order Schmidt modes, initially having higher eigenvalues, become more pronounced.

\subsection{SU(1,1) interferometers}\label{S1.2}

At the very start of nonlinear optics, an idea emerged to realize two nonlinear effects at spatially separated points and to see interference between them. Such \textit{nonlinear interference} would enable the observation of the relative phase between two effects. It was first realized by Chang et al.~\cite{Bloembergen} who measured in this way the complex values of surface quadratic susceptibility for several semiconductors. Later, it became a common way to measure the phases of nonlinear susceptibilities.

\begin{figure}[h]
\begin{center}
\includegraphics[width=0.5\textwidth]{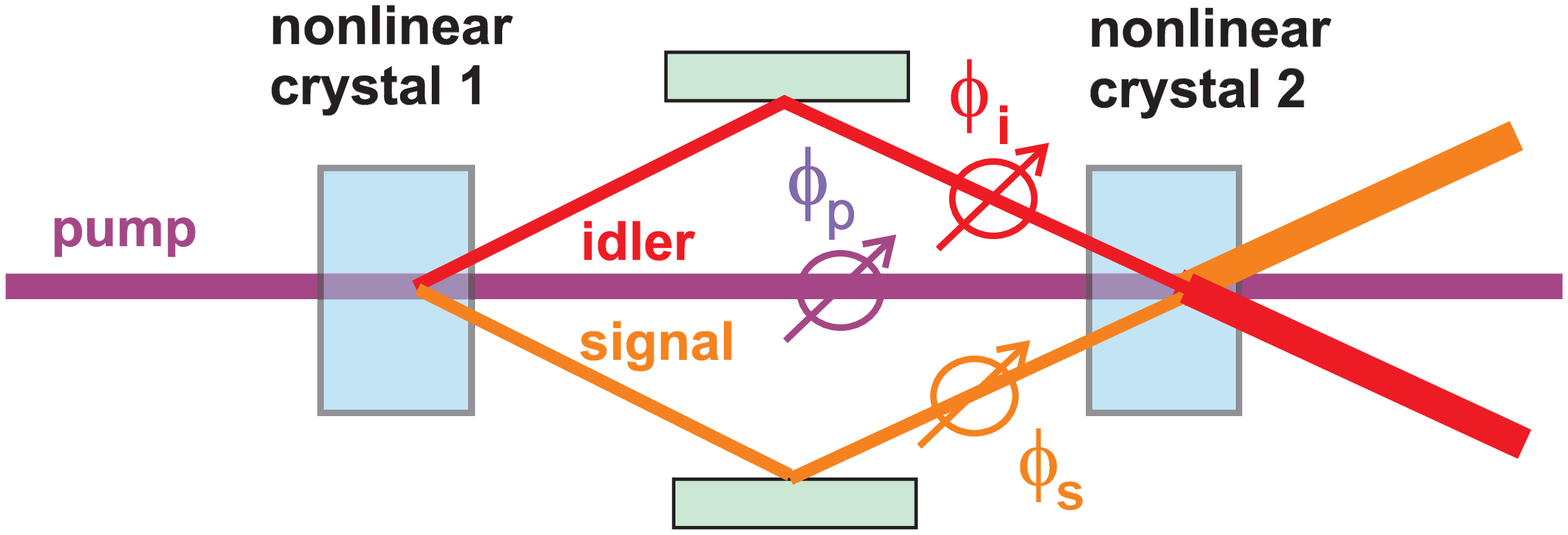}
\caption{(Color online) An SU(1,1) nonlinear interferometer based on two high-gain parametric amplifiers.}\label{nonlinear_int}
\end{center}
\end{figure}

After the discovery of parametric amplification via PDC and four-wave mixing (FWM), it was soon suggested to realize nonlinear interference based on these effects.  Yurke et al.~\cite{SU11} proposed an interferometer in which the signal and idler beams emitted in the first parametric amplifier were directed into the second one and got amplified or deamplified depending on the phases introduced in the pump, signal, or idler beams, $\phi_{p,s,i}$ (Fig.~\ref{nonlinear_int}). Because the transformations performed by the interferometer on the fields at its two output modes relate to the SU(1,1) group, this type of interferometer is usually referred to as SU(1,1).
Initially the SU(1,1) interferometer was proposed as a method to perform phase measurement below the shot-noise level, which is possible due to the strong dependence of its output intensity on the phases $\phi_{s,i}$ at high parametric gain~\cite{Ou}. Nevertheless, the first implementations of SU(1,1) interferometers were based on low-gain (spontaneous) PDC~\cite{Herzog,big}. In particular, in a clever modification of such interferometer the effect of `induced coherence without induced emission' was observed~\cite{induced}, which later was successfully implemented for the measurement of absorption~\cite{abs} and dispersion~\cite{disp}, as well as imaging~\cite{imaging} in the infrared or terahertz~\cite{THz} spectral ranges without the detection of infrared or terahertz radiation.

Only recently, the SU(1,1) interferometer using FWM in rubidium vapor has been implemented for overcoming the shot-noise level of phase measurement~\cite{Zhang}. More than $4$ dB improvement has been obtained compared to a conventional (SU(2)) interferometer populated with the same mean number of photons. The operation was at high gain, which provided $7.4$ dB amplification of the radiation from the first FWM source in the second one.

The same mechanism can be used for shaping the spectrum of high-gain PDC or FWM, both in the angle and in the frequency. Because of the nonlinear amplification of the incident radiation, the modes that are not amplified in the second nonlinear crystal will be much weaker at the output than those amplified. Moreover, one can take advantage of the de-amplification of certain modes, which, however, is not accompanied by a noise increase. Such selective amplification of different modes can enable tailoring the structure of the spectrum.

\section{Diffractive and dispersive spreading, and reduction of the mode number}\label{S2}

\subsection{Angular spectrum tailoring}\label{S2.1}
\begin{figure}[h]
\begin{center}
\includegraphics[width=0.5\textwidth]{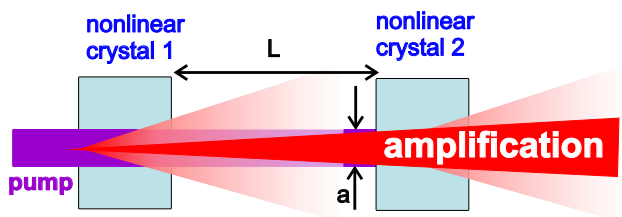}
\caption{(Color online) A nonlinear interferometer formed by two spatially separated crystals.}\label{separated}
\end{center}
\end{figure}

In a nonlinear interferometer formed by two spatially separated traveling-wave high-gain parametric amplifiers (Fig.~\ref{separated})~\cite{separation}, broadband radiation emitted by the first crystal is amplified in the second one. If the distance between the crystals is considerable, only part of the radiation is amplified in the second one - namely, the part that passes through the pump beam in the second crystal. In accordance with this, it was shown~\cite{separation} that at a certain distance between the crystals, the angular spectrum becomes nearly single-mode.
At a sufficiently large distance between the crystals, the angular width of the spectrum amplified in the second crystal should be roughly given by the ratio between the pump diameter $a$ and the distance $L$ between the crystals,
\begin{equation}
\Delta\theta\approx\frac{a}{L}.
\label{eq:angular width}
\end{equation}

This effect of spatial spectrum narrowing has a simple explanation in terms of the Schmidt modes of high-gain PDC. Indeed, each of the spatial Schmidt modes of BSV emitted by the first crystal is diffracting (spreading) in the course of propagation to the second crystal. To a good approximation, the spatial Schmidt modes are given by the Hermite-Gauss or Laguerre-Gauss set, the lowest-order mode being simply a Gaussian beam. Lower-order modes spread slowly and in the second crystal they overlap with the pump beam. Therefore, they get amplified provided that the phase acquired in the course of propagation is appropriate. However, higher-order modes (no matter if they are given by Laguerre-Gauss or Hermite-Gauss beams) spread faster in the space between the crystals and do not get amplified. As a result, the spatial spectrum at the output of the second crystal gets narrower. This continues until only the first Schmidt mode gets amplified, after which the angular width remains constant; increasing the distance $L$ only reduces the total intensity.

The dependence of the angular width $\Delta\theta$ on the distance $L$ between the crystals can be derived from this picture as follows. The zeroth-order Schmidt mode is a Gaussian beam of waist radius $w_{0}$. As it propagates from the crystal, the waist radius at a distance $z$ is~\cite{Kogelnik}
\begin{equation}
w_0(z)=\sqrt{w_{0}^{2}+\left(\frac{\lambda z}{\pi w_{0}}\right)^2},\,\, w_0(0)=w_0,
\label{eq:waist_1st}
\end{equation}
with $\lambda$ being the wavelength. The parameter $\theta_0\equiv\lambda/(\pi w_0)$ is the half-angle divergence of the Gaussian beam. Higher-order modes have larger spatial sizes, $w_{m}=Mw_0$; for instance, for Hermite-Gauss beams, $M=\sqrt{2m+1}$. At the same time, they have larger divergences $\theta_m=M\theta_0$, so that as the distance $z$ increases, they spread as
\begin{equation}
w_m(z)=\sqrt{w_{m}^{2}+\left(M^2\frac{\lambda z}{\pi w_{m}}\right)^2}.
\label{eq:waist_m}
\end{equation}
Assuming that for $z=L$, only modes of orders from $0$ to $m$ are amplified in the second crystal, we find the corresponding $M$ from the condition $w_m(L)=a/2$. We obtain
\begin{equation}
M=\frac{a}{2}\left[w_0^{2}+\left(\frac{\lambda L}{\pi w_0}\right)^2\right]^{-1/2}.
\label{eq:M}
\end{equation}
Then, the divergence of the beam will be equal to twice the half-angle divergence of mode $m$, $\Delta\theta=2\theta_m=M\frac{\lambda}{\pi w_{0}}$:
\begin{equation}
\Delta\theta=\left[\frac{1}{\Delta\theta_0^{2}}+\left(\frac{L}{a}\right)^2\right]^{-1/2},
\label{eq:angular width_full}
\end{equation}
where $\Delta\theta_0=\frac{a\lambda}{\pi w_{0}^2}$ is the initial angular width.

\begin{figure}[h]
\begin{center}
\includegraphics[width=0.4\textwidth]{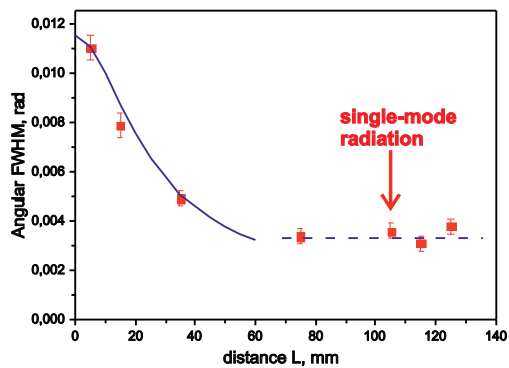}
\caption{(Color online) The angular width of the spectrum measured at the output of the second crystal, versus the distance between the crystals. The solid line is calculated with Eq.~(\ref{eq:angular width_full}), without fitting parameters. The dashed line is a guide to an eye. The arrow shows the position at which $m=1.1$ spatial modes were reported in Ref.~\cite{separation}.}\label{angular}
\end{center}
\end{figure}

We have measured the dependence of the angular width on the distance between the crystals under the same experimental conditions as in Ref.~\cite{separation}: two $3$ mm crystals were placed into a single pump beam of full width at half maximum (FWHM) waist $200$ $\mu$m, and the distance between them was changed from $10$ to $130$ mm. The results are shown in Fig.~\ref{angular}.
Equation (\ref{eq:angular width_full}) was used for fitting the dependence at $L\le60$ mm. At larger distances, the angular spectrum shows no dependence on $L$; the dashed line is just a guide to the eye. The position at which nearly single-mode was observed (the number of modes was measured to be $m=1.1$)~\cite{separation} is shown by an arrow.

\subsection{Frequency spectrum tailoring}\label{S2.2}
\begin{figure}[h]
\begin{center}
\includegraphics[width=0.7\textwidth]{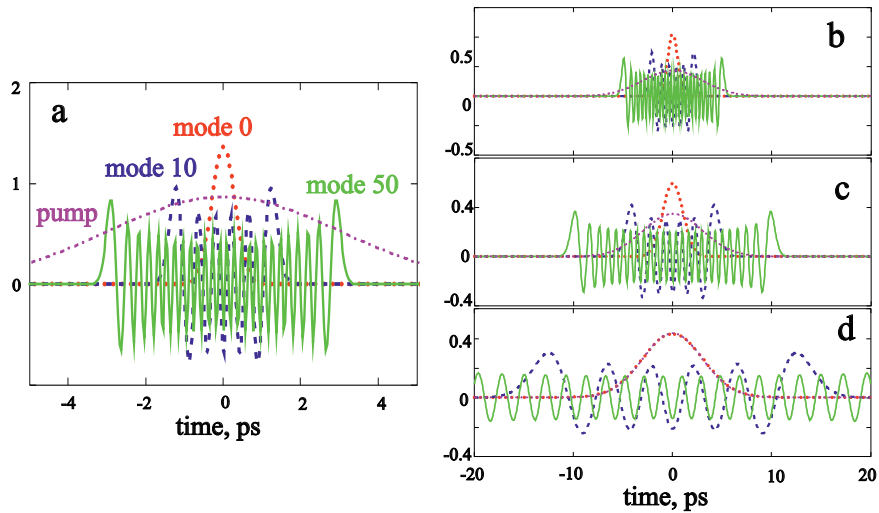}
\caption{(Color online) Temporal Schmidt modes of orders $0$ (red dotted line), $10$ (blue dashed line) and $50$ (green solid line) for a $3$ mm crystal pumped by $6$ ps pulses before (a) and after propagating through $10$ cm (b), $20$ cm (c) and $60$ cm (d) of SF6 glass. For comparison, the amplitude of the pump pulse is shown by magenta dash-dotted line.}\label{modes}
\end{center}
\end{figure}

This effect has an analogue in the frequency/time domain. In this case, the role of the diffractive spreading of beams is played by the dispersive spreading of pulses. Indeed, let a dispersive material of length $d$ be placed between the two crystals. Each frequency Schmidt mode of the BSV from the first crystal will spread in time in the course of propagation through the material, and the spreading will be determined by the group-velocity dispersion (GVD) $k''=d^2k/d\omega^2$. Lower-order modes are narrower in time than higher-order ones, but they will still spread in the GVD material slower than higher-order ones. This is illustrated by Fig.~\ref{modes} where several temporal Schmidt modes are plotted for BSV emitted from a 3 mm crystal pumped by $6$ ps pulses at wavelength $355$ nm (a). The emission is at the degenerate wavelength $710$ nm. The modes are assumed to be the same as for spontaneous parametric down-conversion~\cite{Law}. To a good approximation, they are given by Hermite functions~\cite{Wasilewski}.

After propagation through a dispersive material with the GVD $k''$ and length $d$, the temporal modes will spread in time but maintain their shapes. The latter follows from the fact that, similar to diffractive spreading of beams, dispersive spreading of pulses acts as the Fourier transformation, so that after a sufficiently long GVD material the shape of a pulse becomes similar to its spectral amplitude. At the same time,  Hermite functions are eigenfunctions of the Fourier transformation. Therefore, the whole set of Schmidt modes will be simply rescaled after the propagation through the dispersive material. For instance, the zeroth-order Schmidt mode (dotted line in Fig.~\ref{modes}a), initially a Gaussian pulse of duration $\tau_{0}$ will remain a Gaussian pulse with the duration depending on $d$~\cite{Yariv},
\begin{equation}
\tau_0(d)=\sqrt{\tau_{0}^{2}+\left(\frac{k'' d}{\tau_{0}}\right)^2}.
\label{eq:time_1st}
\end{equation}
Higher-order modes will also maintain their shapes but, similar to the case of the angular modes, will spread faster.

In the right panel of Fig.~\ref{modes}, the Schmidt modes are plotted after the propagation through various lengths $d$ of SF6 glass, whose GVD at the wavelength $710$ nm is $k''=238$ fs$^2$/mm~\cite{refractive index}. In the calculation, Eq.~(\ref{eq:time_1st}) was used for the Gaussian mode, and the higher-order modes were simply rescaled accordingly.) For the length $d=10$ cm (b), only higher-order modes (mode $50$ in the figure) get sufficiently spread so that they do not fully overlap with the pump pulse in the second crystal. Therefore, high-order modes will not be amplified.

 However, at $d=20$ cm (c), even the tenth-order mode becomes considerably spread and will not get fully amplified in the second crystal. This should lead to the narrowing of the spectrum. As the length of the dispersive glass increases, the spectral width should reduce. After $d=60$ cm of glass (d), the zeroth-order Schmidt mode will overlap with the pump pulse. At high gain, this situation should lead to single-mode output emission.

By analogy with Eq.~\ref{eq:angular width_full}, one can estimate the frequency spectrum of BSV generated in the system of two crystals separated by a GVD material as
\begin{equation}
\Delta\omega=\left[\frac{1}{\Delta\omega_0^{2}}+\left(\frac{k''d}{T_p}\right)^2\right]^{-1/2},
\label{eq:frequency width_full}
\end{equation}
where $T_p$ is the pump pulse duration and $\Delta\omega_0$ the initial spectral width.
In the next section, we describe the experimental results confirming this behavior.

\section{Experiment}\label{S3}
\begin{figure}[h]
\begin{center}
\includegraphics[width=0.8\textwidth]{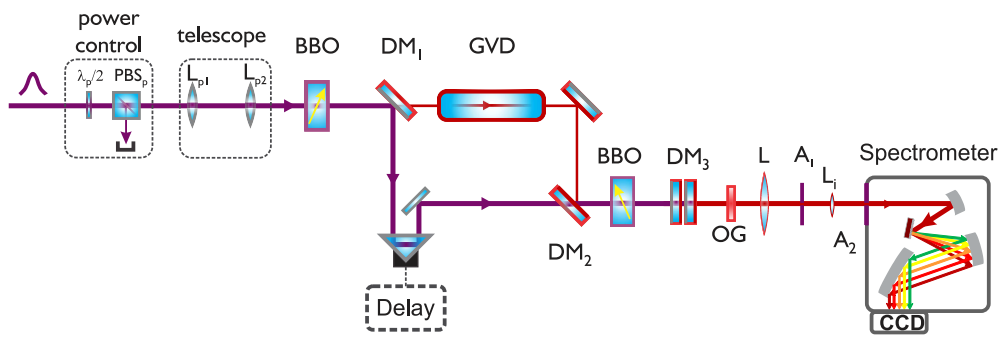}
\caption{(Color online) Schematic of the experimental setup.}\label{setup}
\end{center}
\end{figure}

The scheme of the experimental setup is shown in Fig.~\ref{setup}. Collinear high-gain PDC with the central wavelength at 709.3 nm was created in a type-I 3-mm-long BBO crystal by pumping it with a third harmonic of the pulsed Nd:YAG laser at 354.7 nm, with a pulse duration of 18 ps (the coherence time being $6$ ps) and a repetition rate of 1 kHz. The laser power was varied by a half wave plate  $\lambda_p/2$ followed by a polarization beamsplitter $PBS_p$. A telescope, made of plano convex lenses $L_{p1}$ and $L_{p2}$ with the focal distances of 50 cm and 5 cm, respectively, compressed the beam size down to half-power beam width of 225 $\mu m$. A dichroic mirror $DM_1$ separated the pump beam and the PDC. The PDC pulses were passing through the group velocity dispersion (GVD) medium. We had three options of GVD media: SF-6 glass rods of length $9$ cm and $18.3$ cm, and SF-57 glass rod of length $19.4$ cm. The pump pulses were timed, by means of a delay line, to overlap with the time-stretched PDC pulses on the dichroic mirror $DM_2$ and amplify them in the second type-I 3-mm-long BBO crystal. After the crystal the pump was blocked by a pair of dichroic mirrors $DM_3$ and a long-pass filter OG580. The iris $A_1$ placed in the focal plane of the lens L with the focal distance of 10 cm was used to align the crystals for collinear geometry. The lens $L_i$ focused the PDC radiation onto the input slit of the spectrometer with a resolution of 0.15 nm.

\begin{figure}[h]
\begin{center}
\includegraphics[width=0.7\textwidth]{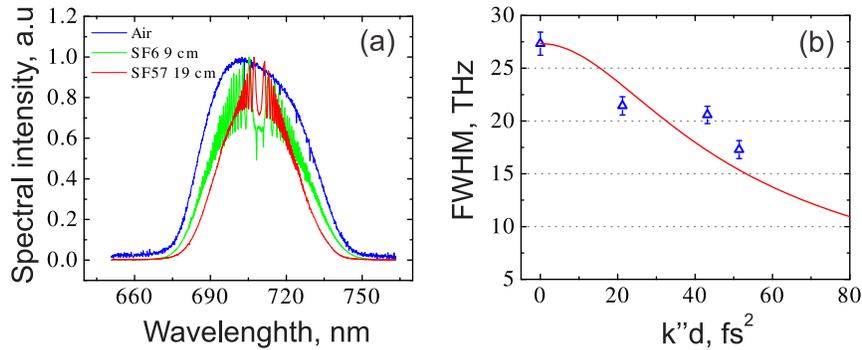}
\caption{(Color online) (a) Measured PDC spectra with different GVD media inserted. (b) FWHM of the PDC spectra experimentally measured with and without GVD media placed between the crystals (blue triangles) and the theoretically calculated dependence according to Eq.~\ref{eq:frequency width_full} (red line) plotted versus
$k''d$. }
\label{results}
\end{center}
\end{figure}

Figure~\ref{results} (a) shows the measured PDC spectra with and without the GVD medium placed between the crystals. The PDC spectrum for the crystals separated by an air gap of $24.2$ cm is shown by a blue line. The measurement was performed at an average pump power of $73.4$ mW in the setup, published before in~\cite{separation}. The interference fringes due to different refractive indexes of the pump and the signal and idler beams in the air were avoided by averaging the spectra taken over different positions of the first crystal in the range from 22.7 cm to 25.7 cm with the step of 2 mm. The FWHM of the spectrum was found to be $45.6$ nm.

Green and red lines in Fig. ~\ref{results} (a) show how the spectrum of the PDC evolves as the value of $k''d$ for the inserted GVD medium is increased. All the spectra are affected by the interference arising from the frequency-dependent phase of the broadband PDC generated in the first crystal after passing through the GVD media. As a result, constructive or destructive interference is observed for the different frequencies at the output of the second crystal. Despite the interference, one can clearly see that the spectra measured with the GVD media are narrower than the one obtained with the air gap between the crystals. The  FWHM of the envelope for each measurement was considered as the width of the spectrum.
In Fig.~\ref{results} (b) we compare the measurement results for the PDC spectral width (blue triangles) and the calculation according to Eq.~\ref{eq:frequency width_full} (red line) using the GVD values for the Schott glass SF6 and SF57~\cite{refractive index}. Instead of the pulse duration $T_p$, the coherence time of the laser $T_c=6 $ ps was used since it is the coherence of the pump that matters for parametric amplification. One can see that the experimental FWHM values agree well with the calculations.

\section{Conclusion}\label{S4}
In conclusion, we have considered a nonlinear interferometer formed by two unseeded traveling-wave parametric amplifiers (based on parametric down-conversion in nonlinear crystals) and showed that its angular and frequency spectrum of emission can be modified by spatially separating the two crystals and/or placing a dispersive material between them. The effect has a simple interpretation in terms of Schmidt modes: higher-order modes spread in space and time faster than low-order ones and hence do not get amplified in the second crystal. Our experimental results show the narrowing of the spatial spectrum, leading ultimately to a single spatial mode. For frequency spectrum narrowing, preliminary results show 30\% narrowing for a glass rod with large group velocity dispersion inserted into the interferometer.

The research leading to these results has received funding from the EU FP7 under grant agreement No. 308803 (project BRISQ2). We thank Xin Jiang and Patricia Schrehardt for providing the samples of SF6 and SF57.

\end{document}